\documentclass[aps,twocolumn,nofootinbib]{revtex4}
\usepackage[latin1]{inputenc}
\usepackage{epsfig}
\usepackage{mathrsfs}
\newcommand{\beq}{\begin{equation}}
\newcommand{\eeq}{\end{equation}}
\newcommand{\la}{\langle}
\newcommand{\ra}{\rangle}

\begin{document}

\title{Derivation of the Schrödinger equation
from classical stochastic dynamics}
\author{Mário J. de Oliveira}
\affiliation{Universidade de São Paulo,
Instituto de Física, Rua do Matão, 1371,
05508-090 São Paulo, SP, Brazil}

\begin{abstract}

From classical stochastic equations of motion we derive
the quantum Schrödinger equation. The derivation
is carried out by assuming that the real and imaginary
parts of the wave function $\phi$ are proportional to the
coordinates and momenta associated to the degrees of
freedom of an underlying classical system. The wave
function $\phi$ is assumed to be a complex time dependent
random variable that obeys a stochastic equation of motion
that preserves the norm of $\phi$. The
quantum Liouville equation is obtained by considering
that the stochastic part of the equation of motion changes
the phase of $\phi$ but not its absolute value. 
The Schrödinger equation follows from the Liouville
equation. The wave function $\psi$ obeying the Schrödinger
equation is related to the stochastic wave function by
$|\psi|^2=\la|\phi|^2\ra$.

\end{abstract}

\maketitle


Schrödinger introduced the quantum wave equation 
that bears his name 
\cite{landau1958,merzbacher1961,messiah1961,
sakurai1967,sakurai1994,griffiths1995,piza2002}
by using an analogy between 
mechanics and optics. Hamilton had shown 
that there is a relation between the principle
of least action of mechanics and the geometric optics.
Considering that the Hamilton equation 
\cite{lanczos1949,goldstein1950,landau1960,arnold1978}
of motion of classical mechanics
is analogous to the equation of geometric optics
then there must be a wave equation, the
Schrödinger equation, which is the analogous
of the light wave equation.

As the theories of motion based on the Hamilton
equation and on the Schrödinger equation describe the
same phenomena, they are conflicting theories as they
predict different results at small scales. However,
at large scales they give the same results and we may
say that in this regime it is possible to derive
the classical Hamilton equation from the Schrödinger
equations. The opposite is never true, what seems
to be in contradiction with the title of this paper.
However, what concerns us here is not the derivation
of quantum mechanics from classical mechanics,
but the derivation of the abstract framework of the
former from that of the latter.

The derivation is accomplished by representing the motion
of a quantum particle by the motion of an {\it underlying}
classical system with many degrees of freedom. 
The coordinate and momentum associated to each degree
of freedom is considered to be proportional to the
real and imaginary parts of a dynamic variable $\phi$.
This relation makes the complex variables $\phi$ and $\phi^*$
a pair of canonically conjugate variables of
the classical underlying system. In other words 
the states of the underlying system is represented
by a phase space with complex components.

The variable $\phi(x)$ is considered to depend on 
a continuous parameter $x$, and is identified as 
the wave function of the quantum system. An observable
is represented by a bilinear functional of $\phi(x)$
and $\phi^*(x)$, considered to be independent variables.
This property of the Hamiltonian makes the norm of
$\phi(x)$ a constant of the motion. This is an
essential property of the underlying system 
which is equivalent to preservation of the inner
product of a quantum state vector. 
As the inner product is preserved, the complex
phase space becomes a Hilbert space and we may
say that the motion of the classical underlying
system is represented in a Hilbert space
\cite{koopman1931,vonneumman1932}.

The probabilistic character of quantum mechanics
is introduced considering the wave function $\phi$
a stochastic variable which means to turn the
equation of motion of the underlying system into
a stochastic equation of motion. This is obtained
by adding a noise to the Hamilton equation of
motion of the underlying system which preserves
the norm of $\phi$ so that the full stochastic
equation of motion preserves the norm.

The underlying system that we consider here is
a classical continuous system in one dimension, 
whose Hamilton equations of motions are
\beq
\frac{\partial q}{\partial t} = \frac{\delta{\cal H}}{\delta p},
\qquad 
\frac{\partial p}{\partial t} = -\frac{\delta{\cal H}}{\delta q},
\eeq
where ${\cal H}$ is the Hamiltonian, which is a functional
of the coordinate $q(x)$ and the canonical momentum $p(x)$,
and we are using a $\delta$ to denote the functional
derivative.

Instead of the pair of real canonical variables $(q,p)$,
we use new variables which are complex variables
obtained through the transformation
$\phi = \alpha q + i \beta p$
and $\phi^* = \alpha q - i \beta p$,
where  $\alpha$ and $\beta$ are real constants
such that $\alpha\beta=1/2\mu$, and $\mu$
is some constant. The new pair $(\phi,\phi^*)$
constitutes a pair of canonically conjugate variables,
in terms of which the equations of motion become
\beq
i \mu \frac{\partial\phi}{\partial t}
= \frac{\delta{\cal H}}{\partial \phi^*},
\qquad
i\mu \frac{\partial\phi^*}{\partial t}
= - \frac{\delta{\cal H}}{\partial \phi},
\label{13}
\eeq 
where $\phi$ and $\phi^*$ are treated as independent
variables, and ${\cal H}$ is now considered a functional
of $\phi(x)$ and $\phi^*(x)$.

Defining the Poisson brackets between two functionals
${\cal A}$ and ${\cal B}$ of $\phi(x)$ and
$\phi^*(x)$ by
\beq
\{{\cal A},{\cal B}\} = \int\left(
\frac{\delta{\cal A}}{\delta\phi}
\frac{\delta{\cal B}}{\delta\phi^*}
- \frac{\delta{\cal B}}{\delta\phi}
\frac{\delta{\cal A}}{\delta\phi^*}\right) dx,
\label{46}
\eeq
then the Hamilton equations can also be written in terms
of Poisson brackets
\beq
i\mu\frac{\partial \phi}{\partial t} = \{\phi,{\cal H}\},
\qquad
i\mu\frac{\partial \phi^*}{\partial t} = \{\phi^*,{\cal H}\}.
\label{11}
\eeq
It is worth pointing out that the canonically
conjugate variables $q$ and $p$ are related by
$\{q,p\} = i\mu$.

The Hamiltonian ${\cal H}$ of the underlying continuous
classical system is assumed to be a bilinear functional
in $\phi(x)$ and $\phi^*(x)$, 
\beq
{\cal H} = \int \phi^*(x)H(x,x')\phi(x') dxdx'.
\label{51}
\eeq
where $H(x,x')=H^*(x',x)$ so that ${\cal H}$ is real.
The norm of $\phi(x)$, defined by
\beq
{\cal N} = \int \phi^*(x)\phi(x) dx,
\label{56}
\eeq
is also understood as a functional of $\phi(x)$ and $\phi^*(x)$.
It is easily seen that ${\cal H}$ of the form
(\ref{51}) commutes in the Poisson sense with ${\cal N}$,
\beq
\{{\cal N},{\cal H}\} = 0,
\eeq
which means that the norm is a constant of the motion.
We choose this constant to be equal to the unity.

If we insert the functional (\ref{51}) into the equation
of motion, either (\ref{13}) or (\ref{11}), we find
\beq
i\mu\frac{\partial\phi}{\partial t} = 
\int H(x,x')\phi(x') dx'.
\label{21}
\eeq

From now on we assume that the wave function $\phi$ is a
a stochastic variable, that is, a time dependent
random variable. It obeys a stochastic dynamics
\cite{kampen1981,risken1989,gardiner2009,tome2015},
which we assume to be the equation of motion (\ref{21}) 
supplemented by a stochastic term,
\beq
\frac{\partial\phi(x)}{\partial t}
= \frac1{i\mu}\int H(x,x')\phi(x')dx' + \zeta(x,t),
\label{21a}
\eeq
where $\zeta(x,t)$ is the stochastic variable 
representing the white noise.

The stochastic variable $\zeta$ is chosen so that the trajectory in 
the vector space spanned by $\phi(x)$ preserves the norm
(\ref{56}). To set up a noise of this type,
we proceed as follows. We discretize the time in
intervals $\tau$ and write the equation (\ref{21a})
in the discretized form
\beq
\delta\phi(x)
= \tau f(x) + i\sqrt{\tau} g(x)\xi - \frac12 \tau k(x),
\label{37}
\eeq
where $\xi$ is a random variable with zero mean and variance
equal to the unity, and $f$, $g$, and $k$ are functions of
$x$, given by
\beq
f = \frac1{i\mu} \int H(x,x')\phi(x') dx',
\eeq
\beq
g = \int G(x,x') \phi(x')dx',
\eeq
\beq
k = \int K(x,x') \phi(x')dx'.
\eeq
The increment in the norm due to an increment in
$\phi$ is
\[
\delta {\cal N} = \tau \int (f \phi^* + \phi f^*)dx
+ i \sqrt{\tau} \xi \int(g \phi^* - \phi g^*)dx 
\]
\beq
+ \frac12\tau\int(2g g^* - k \phi^* - \phi k^* )dx.
\eeq
The first term vanishes identically, and
the second term equals
\beq
i\sqrt{\tau}\xi\int [(G(x',x)
-  G^*(x,x')]\phi(x) \phi(x')^*dx'dx.
\eeq
The last term vanishes if we choose
$K$ to be related to $G$ by
\beq
K(x,x') = \int G(y,x')G^*(y,x)dy.
\eeq

If we choose $G(x',x)=G^*(x,x')$ then $\delta{\cal N}$
vanishes identically and ${\cal N}$ will be invariant
along a trajectory in the vector space in spite of
the trajectory being stochastic. If this condition is
not satisfied, $\delta{\cal N}$ will vanish in the
average and $\la{\cal N}\ra$ will be constant
which we choose to be equal to unity.

Let us determine the increment in $\phi(x)\phi^*(x')$.
Up to terms of order $\tau$ it is given by
\[
\delta(\phi_1\phi_2^*)
= \tau (f_1 \phi_2^* + \phi_1 f_2^*)
+ \sqrt{\tau} \xi(g_1 \phi_2^* + \phi_1 g_2^*) 
\]
\beq
+ \tau  (k_1 \phi_2^* + \phi_1 k_2^* + g_1 g_2^*),
\eeq
where we are using the index 1 and 2 do denote
functions of $x$ and $x'$, respectively.
Taking the average of this equation, the 
second term proportional to $\sqrt{\tau}$
vanishes. Dividing the result by $\tau$ we find
the equation for the time evolution of 
the covariances
$\rho(x,x')=\la \phi(x)\phi^*(x')\ra$,
\beq
\frac{\partial\rho(x,x')}{\partial t}
= \la f_1 \phi_2^* + \phi_1 f_2^*\ra
+ \frac12\la 2g_1 g_2^*- k_1 \phi_2^* - \phi_1 k_2^*\ra.
\eeq
Replacing the expressions of $f$, $g$ and $k$
in this equation we find
\[
\frac{\partial\rho(x,x')}{\partial t} =
\frac1{i\mu} \int[H(x,y)\rho(y,x') -\rho(x,y) H(y,x')] dy
\]
\[
+ \int G(x,y)\rho(y,y') G^*(x',y')  dy dy'
\]
\[
-\frac12 \int G^*(y',x) G(y',y)\rho(y,x')dydy'
\]
\beq
-\frac12\int \rho(x,y) G^*(y',y)G(y',x')dy dy'.
\label{55}
\eeq
As the norm is preserved in the average it follows
from (\ref{56}) that
\beq
\int \rho(x,x)dx = 1.
\label{55a}
\eeq

The equation (\ref{55}) is a closed equation for 
the covariances $\rho(x,x')=\la\phi(x)\phi^*(x')\ra$
because high-order correlations are not involved.
Equations for these high-oder correlations could
also be obtained. However this is not necessary if
we wich to determine the averages of bilinear
functionals such as 
\beq
{\cal A} = \int A(x,x')\phi(x')\phi^*(x)dx dx'.
\label{36}
\eeq
Its average is obtained from $\rho(x',x)$ by
\beq
\la{\cal A}\ra = \int A(x,x')\rho(x',x)dx dx'.
\label{57}
\eeq

Equation (\ref{55}) is the fundamental equation that
we wished to derive. From this fundamental equation
we obtain the quantum Liouville equation and the
Schrödinger equation by choosing a noise that changes
the phase of $\phi(x)$ but not its absolute value.
This is accomplished by choosing
$G(x,x')=\gamma \delta(x-x')$.
In this case the terms of the equation (\ref{55})
involving $G$ vanish and the equation is reduced
to the equation
\beq
i\mu\frac{\partial}{\partial t}\rho(x,x')
= \int[H(x,y)\rho(y,x') -\rho(x,y) H(y,x')] dy,
\label{44}
\eeq
which is the quantum Liouville equation,
if we set $\mu=\hbar$. 

It is easily seen that the quantum Liouville equation
(\ref{44}) admits solutions of the type
\beq
\rho(x,x')=\psi(x)\psi^*(x').
\label{24}
\eeq
called pure states.
If we replace this expression in (\ref{44}), we found that
this form is indeed a solution as long as $\psi(x)$
obeys the equation 
\beq
i\mu \frac{\partial}{\partial t}\psi(x)
= \int H(x,x')\psi(x')dx',
\label{38a}
\eeq
which is identified as the Schrödinger equation,
if we set $\mu=\hbar$. In other words, the Schrödinger
equation is a particular case of the quantum Liouville
equation (\ref{44}) which is obtained when the initial
condition is of the pure state type 
$\rho_0(x,x')=\psi_0(x)\psi_0^*(x')$,
if this is allowed by the physical conditions.
To avoid confusion with the wave function $\phi(x)$,
which is a stochastic variable, we call $\psi(x)$
the Schrödinger wave function.

It is worthwhile writing down the equation 
that gives the time evolution of the average
$\sigma(x)=\la\phi(x)\ra$ of the wave function
$\phi(x)$. It is obtained by taking the average
of (\ref{37}) and the result is
\beq
\frac{d\sigma(x)}{dt} = \frac1{i\mu}\int H(x,y)\sigma(y) dy
- \frac12 \int K(x,y)\sigma(y) dy
\eeq
If the noise changes only the phase of $\phi(x)$,
then $K(x,x')=\gamma^2 \delta(x-x')$ and
\beq
\frac{d\sigma(x)}{dt} = \frac1{i\mu}\int H(x,y)\sigma(y) dy
- \frac12 \gamma^2 \sigma(x).
\label{47}
\eeq
We remark that in the long run, $\sigma$ vanishes and
cannot be identified with the Schrödinger wave function $\psi$,
which obeys the Schrödinger equation (\ref{38a}). 

We point out that in accordance with the present
approach the equation (\ref{55}) as well as the
quantum Liouville equation (\ref{44}) and the Schrödinger
equation (\ref{38a}) that follows from (\ref{55}) are
equations for the covariances
$\rho(x,x')=\la\phi(x)\phi(x')\ra$ of the
stochastic wave equation $\phi(x)$. In particular
$\rho(x,x)=\la|\phi(x)|^2\ra$ is the variance of
the stochastic variable $\phi(x)$ which in the
case of pure states is
\beq
|\psi(x)|^2=\la|\phi(x)|^2\ra
\eeq
and, in view of the normalization (\ref{55a}), obeys
the normalization
\beq
\int |\psi(x)|^2 dx =1.
\label{58}
\eeq

In accordance with the present approach the variable
$x$ is a continuous parameter of the wave function and
$|\psi(x)|^2$ is the covariance of the wave function.
The variable $x$ is not properly a random variable.
However, in view of the normalization (\ref{58}),
$|\psi(x)|^2$ is understood in the usual interpretation
of quantum mechanics as the probability density
distribution of $x$.

The equations that we have derived can be written in a
more compact and familiar form by defining the operators
associated to a bilinear functional such as that given
by (\ref{36}). We define the operator $\hat{A}$ acting
on the vector space by
\beq
\hat{A}\phi(x) = \int A(x,x')\phi(x') dx'.
\eeq
so that the bilinear functional ${\cal A}$ is
\beq
{\cal A} = \int \phi^*(x) \hat{A}\phi(x) dx
\label{30}
\eeq
In terms of the operators $\hat{H}$, $\hat{\rho}$
and $\hat{G}$ associated to $H(x,x')$, $\rho(x,x')$,
and $G(x,x')$, the equation (\ref{55}) acquires the form
\beq
\frac{\partial\hat{\rho}}{\partial t} = \frac1{i\hbar}[\hat{H},\hat{\rho}]
+ \hat{G}\hat{\rho}\,\hat{G}^\dagger
- \frac12\hat{G}^\dagger \hat{G}\hat{\rho}
- \frac12\hat{\rho}\,\hat{G}^\dagger\hat{G},
\label{45}
\eeq
which has the form of the Lindblad equation for open quantum
systems \cite{lindblad1976,breuer2002}.
We point out that ${\rm Tr}\hat{\rho}=1$, which
follows from (\ref{55a}), and that
$\la{\cal A}\ra = {\rm Tr} \hat{A}\hat{\rho}$,
which follows from (\ref{57}). The quantum
Liouville equation (\ref{44}) and the Schrödinger
equation (\ref{38a}) acquire the familiar forms
\beq
i\hbar\frac{\partial\hat{\rho}}{\partial t}
= [\hat{H},\hat{\rho}]
\eeq
and
\beq
i\hbar\frac{\partial\psi(x)}{\partial t} = \hat{H}\psi(x),
\label{70}
\eeq
In the case of pure states, that are solutions of
the Schrödinger equation, the average $\la{\cal A}\ra$
is reduced to 
\beq
\la{\cal A}\ra = \int \psi^*(x) \hat{A}\psi(x) dx
\eeq
which is the familiar form for the quantum average of
an observable.


The Schrödinger equation (\ref{38a}) is formally 
identical to the equation (\ref{21}) and thus
the Schrödinger wave function $\psi(x)$ can 
be understood as related to the coordinate
and momentum of the underlying classical system
described by the classical Hamiltonian (\ref{51})
which is now written in terms of $\psi$ as 
\beq
{\cal H} = \int \psi^*(x)H(x,x')\psi(x') dxdx'.
\label{51a}
\eeq

The question that now arises is how to describe
the position and momentum of the quantum particle
in terms of the coordinates and momenta of the
underlying classical system, which are the real
part and imaginary parts of $\psi$, respectively.

We start by assuming that the position ${\cal X}$
of the quantum particles is given by 
\beq
{\cal X} = \int \psi^*(x) x\psi(x) dx.
\label{42}
\eeq
In accordance with (\ref{30}), the corresponding
operator $\hat{x}$ is such that $\hat{x}\psi(x)=x\psi(x)$.
Also in accordance with (\ref{30}), the momentum ${\cal P}$
is of the form
\beq
{\cal P} = \int \psi^*(x) \hat{p}\psi(x) dx.
\label{43}
\eeq

To determine the momentum ${\cal P}$, we regard it
as the canonical conjugate to the position
${\cal X}$ and should then be related by 
the Poisson commutation of the type $\{q,p\} = i\mu$,
that is, $\{{\cal X},{\cal P}\} = i\mu$.
Replacing ${\cal X}$ and ${\cal P}$ given by
(\ref{42}) and (\ref{43}) in the expression (\ref{46}) for
the Poisson brackets we find
\beq
\{{\cal X},{\cal P}\} = 
\int \psi^*(x) [\hat{x},\hat{p}] \psi(x) dx.
\eeq 
Therefore, the following expression
\beq
\int (\psi^*(x) x\hat{p}\psi(x) - \psi^*(x)\hat{p}x\psi(x) dx
=i\mu
\eeq 
should be fulfilled for an arbitrary function $\psi(x)$.
This is accomplished if we choose $\hat{p}$ 
to be the differential operator
\beq
\hat{p}\,\psi(x)= -i\mu\frac{\partial}{\partial x}\psi(x).
\label{39}
\eeq

The Hamiltonian ${\cal H} = {\cal K}+ {\cal V}$ is
a sum of the kinetic energy ${\cal K}$ plus the
potential energy ${\cal V}$, which is expressed by
\beq
{\cal V} = \int \psi^*(x) V(x) \psi(x) dx.
\eeq
To find the expression for the kinetic energy
we used the relation
$\{{\cal X},{\cal K}\} = {\cal P}/m$, where $m$
is the mass of the quantum particle,
from which follows that the respective operators
are related by $[\hat{x},\hat{K}] = \hat{p}/m$.
Therefore, we may conclude that $\hat{K}=\hat{p}^2/2m$,
and that
\beq
\hat{H} = \frac{\hat{p}^2}{2m} + V.
\eeq
Replacing this result in (\ref{70}) and taking into
account that $\hat{p}$ is given by (\ref{39}), we
reach the equation
\beq
i\mu\frac{\partial\psi}{\partial t} = 
- \frac{\mu^2}{2m}\frac{\partial^2\psi}{\partial x^2}
+ V\psi,
\eeq
which is the original equation introduced by Schrödinger.

In summary, we have derived the Schrödinger equation
from the equations of motion of a
classical underlying system with many degrees of
freedom. The classical system is represented in
a complex phase space whose components which are
the dynamic variables are identified as the wave
function $\phi(x)$. This complex phase space turns out to
be a Hilbert space since the motion of the underlying
classical system preserves the norm of $\phi(x)$.
The probabilistic character of quantum mechanics
is introduced by turning the wave function $\phi$
a stochastic variable, which is accomplished by
adding a noise the the Hamilton equations,
that preserves the the norm of $\phi(x)$.

From these assumptions we derived a general
equation for the covariances $\rho(x,x')$ of
$\phi(x)$ that turns out to be of the Lindblad form.
When the noise changes the phase of $\phi(x)$ but
not its absolute value, this equation turns
out to be the quantum Liouville equation for
$\rho(x,x')$. A special type of solution of
the quantum Liouville equation is of the
type $\rho(x,x')=\psi(x')^*\phi(x)$.
When this happens, $\psi(x)$ obeys an equation
which turns out to be the Schrödinger equation.

The present approach to the Schrödinger equation
does not reduce the science of quantum mechanics
into the science of classical mechanics as
the underlying classical system is not an observable.
But the present approach allows for an an
interpretation of the wave function different from the
standard interpretation
\cite{omnes1994,auletta2001,freire2022}.
Taking into account that
the classical underlying system is a collection of
interacting harmonic oscillator, the motion of the
system can be understood as a wave motion, fitting
the de Broglie ideas about quantum motion \cite{debroglie1927}.
As the trajectory in the Hilbert space is stochastic,
enabling several trajectories,
it may also fit the consistent histories interpretation
of quantum mechanics \cite{griffiths1995}.


\end{document}